\documentstyle[12pt]{article}
\newcommand{\be}{\begin{eqnarray}}
\newcommand{\ee}{\end{eqnarray}}
\def\rgh{\raise0.3ex\hbox{$\longrightarrow$\kern-0.75em\raise-1.1ex\hbox
{$\scriptstyle q\rightarrow i$}}}
\begin{document}
\begin{titlepage}
\hspace*{10.0cm}{\bf HU-TFT-93-23}
\vskip 1.0 cm
\begin{center}
{\large\bf Statistics of q-Oscillators, Quons and Relations\\
to Fractional Statistics}
\vskip 2.0 cm
by
\vskip 0.5 cm

{\bf M. Chaichian$^1$, R. Gonzalez Felipe$^{1,*}$\renewcommand{\thefootnote}{\
}
\footnote{$^{*)}$ \small ICSC-World Laboratory; On leave of absence from
Grupo de Fisica Te\'orica, Instituto de
Cibern\'etica, Matem\'atica y Fisica, Academia de Ciencias de Cuba, Calle
E No. 309, Vedado, La Habana 4, Cuba.}and C. Montonen$^2$}
\vskip 3.0 cm
\parbox[t]{12.0cm}{$^1$High Energy Physics Laboratory, Department of Physics,\\
P.O. Box 9 (Siltavuorenpenger 20 C), SF-00014 University of Helsinki, Finland}
\vskip 0.5 cm
\parbox[t]{12.0cm}{$^2$Department of Theoretical Physics, P.O. Box 9
(Siltavuorenpenger 20 C), SF-00014 University of Helsinki, Finland}
\vskip 0.5cm

{\bf March 1993}

\end{center}
\vskip 1.5 cm
Abstract:

The statistics of $q$-oscillators, quons and to some extent, of anyons are
studied and the basic differences among these objects are pointed out. In
particular, the statistical distributions for different bosonic and fermionic
$q$-oscillators are found for their corresponding Fock space representations
in the case when the hamiltonian is identified with the number operator. In
this case and for nonrelativistic particles, the single-particle temperature
Green function is defined with $q$-deformed periodicity conditions. The
equations of state for nonrelativistic and ultrarelativistic bosonic $q$-gases
in an arbitrary space dimension are found near Bose
statistics, as well as the one for an anyonic gas near Bose and Fermi
statistics. The first corrections to the second virial coefficients are
also evaluated. The phenomenon of Bose-Einstein condensation in the
$q$-deformed
gases is also discussed.
\end{titlepage}

\section{Introduction.}

In the last few years the interest in particles obeying statistics different
from Bose or Fermi has increased considerably. The observation in 1953 by
Green \cite{green} that different kinds of statistics are allowed within the
context of quantum field theory (QFT) has stimulated an extensive study of the
so-called paraBose and paraFermi statistical fields \cite{grecra}. The
commutation relations for these fields are trilinear in the creation and
annihilation operators and are characterized by an integer $p$, the order of
the parastatistics, which corresponds to the number of particles in a given
symmetric or antisymmetric state. (If $p=1$, one recovers the usual Bose and
Fermi statistics). The case when $p$ is not an integer has been recently also
studied \cite{igna} in order to provide theories in which the Pauli exclusion
principle and/or Bose statistics can be slightly violated. However, the
corresponding QFTs, which describe such particles, turn out to have negative
norm states and consequently are not physically acceptable. More recently,
another possibility has been explored, which corresponds to the case where
no assumption is made about the parameter $p$ \cite{greenberg}-\cite{alth}.
The particles which obey this type of statistics are called "quons" and the
statistics is referred to as "infinite statistics". They can be realized by
different commutation relations like the ones of $q$-deformed algebras.

On the other hand quantum groups \cite{faddeev} are a subject of great activity
at present and although their direct physical interpretation is still lacking,
it is of particular importance to study the possible physical implications of
these deformations. These structures which first emerged in connection with
the quantum inverse scattering theory \cite{sklya} and solvable statistical
mechanical models \cite{kulish}, have found recent important
applications in many problems of physical and mathematical interest, such as
rational conformal field theories \cite{alvarez}, non-commutative geometry
\cite{manin}, knot theory \cite{kauffman}, quantum superalgebras \cite{chaic}
and so on. The concept of $q$-deformed oscillators (the so-called
$q$-oscillators), \cite{mac} derived from the contraction of quantum algebra
\cite{chaic}, has also been introduced. These $q$-oscillators can formally
be defined in any space dimension but then they violate the fundamental
axioms of QFT on the relation between spin and statistics. In two dimensions,
however, a new kind of statistics interpolating between bosons and fermions
can (without violating any of the axioms) exist, due to the appearance of the
braid group instead of the permutation group when identical particles are
exchanged. The particles which obey such so-called fractional statistics are
called anyons \cite{wile}. Such particles have attracted considerable interest
since they appear in realistic systems and in particular, are connected with
the interpretation of the fractional quantum Hall effect \cite{quant} and may
have relevance in other quasi-two-dimensional systems such as high $T_c$
superconductors \cite{laug}.

Since anyons are two-dimensional particles acquiring a phase upon circling
each other and returning to their original configuration, their wavefunctions
as well as the operators that create or annihilate them must be multi-valuated
functions of position. This fact makes the construction of such operators
rather involved. This problem was originally studied for field theories in the
continuum \cite{seme},\cite{jackiw} but there the precise definition of
multi-valued operators encounters severe difficulties. The same problem,
however, has been lately examined for  field theories on a lattice \cite{frad},
\cite{lerda}, where the multi-valued anyon operators can be indeed rigorously
defined.

In this paper we would like to study the interrelation between the statistics
of $q$-oscillators, quons and, partially, of anyons. At the same time we point
out the basic differences among the statistics of the above mentioned objects.
Along this line we hope to be able to use the techniques, such as the Green
function method developed here, the Fock space representation, etc., for the
$q$-oscillators, further in the case of fractional statistics.

In sect. 2, the $q$-deformed bosonic and fermionic oscillators are defined
and their Fock space representations are given for different forms of their
commutation relations. We emphasize the basic differences between
$q$-oscillators, quons and anyons in general. Sect. 3 is devoted to the study
of the statistics of $q$-oscillators. In particular, the statistical
"distributions" for $q$-bosons and $q$-fermions are found in the case when the
hamiltonian of the system is identified with the number operator. Comments on
the results previously obtained in the literature, are presented. In sect. 4
the
$q$-deformed single-particle temperature Green function is defined and its
"periodicity" properties are studied in order to get an appropriate Fourier
decomposition in the Euclidean "time" variable and find the corresponding
frequency values. The method for evaluating frequency sums is also discussed
and in particular, is used to calculate the statistical "distribution" for
$q$-particles. In sect. 5 we consider an ideal $q$-gas described by a general
noninteracting hamiltonian and find perturbatively the equation of state near
Bose statistics in different cases, namely, nonrelativistic and
ultrarelativistic $q$-boson gas in an arbitrary space dimension.
The low-density regime is also considered to find the virial expansion and the
corrections to the second virial coefficient. Sect. 6 is devoted to the study
of a gas of anyons near Bose and Fermi statistics. We find perturbatively, in
the first order of the statistics determining parameter, the equation of
state of the anyonic gas by means of the temperature Green function approach.
Finally, concluding remarks and directions for further studies are given
in sect. 7.
\vskip 1.0 cm
\section{$q$-oscillator algebras.}

In this section we shall review briefly the definition of $q$-deformed bosonic
and fermionic oscillators and their Fock space representations. As in the
classical case, the $q$-oscillators can be obtained from the $SU(2)_q$ algebra
by a contraction procedure \cite{chaic}.

We start by defining the $q$-bosonic oscillator algebra through the commutation
relations \cite{mac},\cite{chaic}
$$
aa^+-qa^+a=q^{-N}\ ,$$
$$[N,a]=-a,\ [N,a^+]=a^+\ ,
\eqno{(2.1)}
$$
where $q\in{\bf C}\ ;\ a,a^+$ and $N$ are the annihilation, creation and number
operators respectively. One can construct the representation of (2.1) in the
Fock space spanned by the orthonormalized eigenstates $|n>$ of the operator
$N$,
$$
\renewcommand{\theequation}{2.\arabic{equation}}
|n>=\frac{(a^+)^n}{\sqrt{[n]!}}|0>\;,\ \ a|0>=0\;,\ \ N|n>=n|n>\;,
\eqno{(2.2)}
$$
where $[n]!\equiv [n][n-1]...[1],[0]!=1$ and $[n]$ denotes
$$
\renewcommand{\theequation}{3.\arabic{equation}}
[n]\equiv\frac{q^n-q^{-n}}{q-q^{-1}}\ .
\eqno{(2.3)}
$$
In this Fock space it is easy to prove that the following relations hold:
$$
\renewcommand{\theequation}{4.\arabic{equation}}
a^+a=[N],\ \ aa^+=[N+1]\ .
\eqno{(2.4)}
$$
It is interesting to note that if $q=e^{\pm i\pi/m}$, then $[m]=0$ (see
definition (2.3)). Consequently, the Fock space breaks up into $m$-dimensional
subspaces not connected by the operators $a$ and $a^+$. Each subspace carries
an $m$-dimensional representation of the algebra (2.1) and can be
considered separately. (An example is given below, eq. (3.6)). A corresponding
phenomenon happens whenever $q=\exp (ir\pi)$, where $r$ is rational, and
different from an integer.

These exist different forms of relations (2.1). In particular, when $q$ is
real, it is possible to define the operators \cite{kuda}
$$
A=q^{\frac{N}{2}}a\ ,\ A^+=a^+q^{\frac{N}{2}}\ ,
\eqno{(2.5)}
$$
which satisfy the commutation relation
$$
AA^+-q^2A^+A=1\ .
\eqno{(2.6)}
$$
In terms of the new operators (2.5), the Fock representation (2.2) reads as
$$
|n>=\frac{(A^+)^n}{\sqrt{[n]^B!}}|0>\;,\ A|0>=0\;,\ N|n>=n|n>\;,
\eqno{(2.7)}
$$
where
$$
[n]^B\equiv q^{n-1}[n]=\frac{q^{2n}-1}{q^2-1}\ .
\eqno{(2.8)}
$$
(B standing for bosons).

\noindent Finally in this Fock space we have
$$
A^+A=[N]^B\;,\ AA^+=[N+1]^B\ .
\eqno{(2.9)}
$$
One can also express (2.1) in terms of the usual undeformed Bose operators
$b,b^+$ by means of the change \cite{kuda}
$$
a=\left(\frac{[N+1]}{N+1}\right)^{1/2}b\;,\ \ a^+=b^+\left(\frac{[N+1]}{N+1}
\right)^{1/2}\ .
\eqno{(2.10)}
$$
Then we obtain $[b,b^+]=1,\ [N,b]=-b,\ [N,b^+]=b^+,\ N=b^+b$, i.e., the usual
bosonic oscillator algebra.

Let us now introduce the fermionic $q$-oscillator algebra. The annihilation and
creation operators of the fermionic $q$-oscillator $f,f^+$ are postulated to
satisfy the commutation relations \cite{parth}
$$
ff^+ +qf^+f=q^{-N}\ ,
\eqno{(2.11)}
$$
$$
[N,f]=-f\;,\ [N,f^+]=f^+\ ,
$$
where $N$ is the $q$-fermion number operator. The orthonormalized eigenstates
of $N$ are defined by
$$
|n>=\frac{(f^+)^n}{\sqrt{[n]^f!}}|0>\;,\ f|0>=0\;,\ N|n>=n|n>\ ,
\eqno{(2.12)}
$$
where
$$
[n]^f\equiv \frac{q^{-n}-(-1)^nq^n}{q+q^{-1}}\ ,
\eqno{(2.13)}
$$
and we have
$$
f^+f=[N]^f\;,\ ff^+=[N+1]^f\ .
\eqno{(2.14)}
$$
For generic $q$, this representation is infinite-dimensional. Like in the
$q$-bosonic case, for special values of $q$ the Fock space breaks up into
disjoint subspaces each carrying a finite-dimensional representation of (2.11).
If, in particular $q=e^{\pm i\pi/m}$, we have $[2m]^f=0$ for $m$ odd, whereas
for $m=4k,\ [m]^f=0$ and for $m=4k+2,\ k\geq 1\ [\frac{m}{2}]^f=0$; the
dimensions of the disjoint subspaces being the smallest integer $n$ for which
$[n]^f=0$. (The cases $q=e^{\pm i\pi/2}$ are special, with e.g. $[n]^f
\rgh -i^{-(n+1)}n$; they thus give infinite-dimensional
representations). For $q=1$ the Fock space breaks up into two-dimensional
subspaces, and the Pauli exclusion principle follows from $f^2=(f^+)^2=0$.

Let us introduce now the modified operators
$$
F=q^{\frac{N}{2}}f\;,\ F^+=f^+q^{\frac{N}{2}}\ ,
\eqno{(2.15)}
$$
for $q$ real, which satisfy the commutation relation
$$
FF^+ +q^2F^+F=1\ .
\eqno{(2.16)}
$$
Using (2.12) one can build the Fock space representation in terms of these new
operators. We obtain
$$
|n>=\frac{(F^+)^n}{\sqrt{[n]^{F}!}}|0>\;,\ F|0>=0\;,\ N|n>=n|n>\ ,
\eqno{(2.17)}
$$
with
$$
[n]^F\equiv q^{n-1}[n]^f=\frac{1-(-1)^nq^{2n}}{q^2+1}\ .
\eqno{(2.18)}
$$
It is worthwhile to remark that in the case of fermionic $q$-oscillators
defined
through the commutation relations (2.11), there does not exist a change of
operators, analogous to (2.10), which would allow us to express these
oscillators in terms of the usual (undeformed) Fermi ones. One can use instead
another definition for fermionic $q$-oscillators \cite{chaic},
$$
cc^++qc^+c=q^M\ ,$$
$$[M,c]=-c\;,\ [M,c^+]=c^+\ .
\eqno{(2.19)}
$$
In this case the change of operators
$$
C=q^{-\frac{M}{2}}c\ ,\ C^+=c^+q^{-\frac{M}{2}}
\eqno{(2.20)}
$$
would lead to the usual Fermi commutation relations
$$CC^+ +C^+C=1\ ,
$$
and consequently, the only non-vanishing eigenstates of the number operator
$M$ are $|0>$ and $|1>=C^+|0>$. (Notice that this conclusion derived from
(2.19) is equivalent to the relations $C^2=C^{+^2}=0$ and thus, there is no
need to assume them). This means that fermionic $q$-oscillators defined by
the algebra (2.19) have only the trivial (undeformed) Fock space representation
and thus, they behave as usual fermions.

To conclude this section, some remarks about systems with several degrees of
freedom are in order. In the case of bosonic and fermionic $q$-oscillator
algebras (2.1), (2.5) and (2.11), (2.15) respectively, the multimode
generalization is straightforward. For $q$-bosons we have the commutation
relations
$$
a_ia^+_j-((q-1)\delta_{ij}+1)a^+_ja_i=\delta_{ij}q^{-N_i}\ ,$$
$$[N_i,a_j]=-\delta_{ij}a_j\ ,\ [N_i,a^+_j]=\delta_{ij}a^+_j\ ;
\eqno{(2.21)}$$

$$A_iA^+_j-((q^2-1)\delta_{ij}+1)A^+_jA_i=\delta_{ij}\ ,$$
$$[N_i,A_j]=-\delta_{ij}A_j\ ,\ [N_i,A^+_j]=\delta_{ij}A^+_j\ ;
\eqno{(2.22)}
$$
while for $q$-fermions they read as
$$f_if^+_j+((q-1)\delta_{ij}+1)f^+_jf_i=\delta_{ij}q^{-N_i}\ ,$$
$$[N_i,f_j]=-\delta_{ij}f_j\ ,\ [N_i,f^+_j]=\delta_{ij}f^+_j\ ;
\eqno{(2.23)}
$$

$$F_iF^+_j+((q^2-1)\delta_{ij}+1)F^+_jF_i=\delta_{ij}\ ,$$
$$[N_i,F_j]=-\delta_{ij}F_j\ ,\ [N_i,F^+_j]=\delta_{ij}F^+_j\ .
\eqno{(2.24)}
$$
Recently, an alternative deformed commutation relation, called quon algebra,
has been considered \cite{greenberg},\cite{moha}. It is defined by
$$
a_ia^+_j-qa^+_ja_i=\delta_{ij}\ ,
\eqno{(2.25)}
$$
and interpolates between Bose and Fermi algebras as $q$ goes from 1 to -1
on the real axis. However, let us note that there exists a distinct
difference between the $q$-oscillators defined in (2.22), (2.24) and the
quons. In fact, in the case of bosonic (fermionic) $q$-oscillators different
modes $(i\neq j)$ commute (anticommute), while quons "$q$-mute" i.e. satisfy
the relation given in (2.25). Moreover, in the case of quon algebra no
commutation rule can be imposed on $aa$ and $a^+a^+$. In effect, relations
such as $a_ia_j-qa_ja_i=0$ or $a^+_ia^+_j-qa^+_ja^+_i=0$ would hold only
when $q^2=1$, i.e., only in the Bose and Fermi cases.

As far as anyons are concerned, their braiding properties make the construction
of an oscillator algebra a difficult and still unsolved problem. In a recent
paper the construction of anyonic oscillators on a lattice as well as their
relation with quantum groups have been studied \cite{lerda}. We will not
discuss them further here, but it is worthwhile to stress that $q$-oscillators
and anyonic oscillators differ substantially from each other. First of all,
anyons arise only in two dimensions, while $q$-oscillators can be defined in
any space dimension. Secondly, as we shall see in section 4, the
$q$-oscillators
can be interpreted as the Fourier components of local field operators,
whereas anyons are non-local objects as a consequence of their braiding
properties and the essential differences in their commutation relations.
\vskip 1.0 cm
\section{$q$-deformed statistics.}

In order to study the properties of $q$-oscillators, let us consider now the
statistical averages and calculate the deformed "distributions" which follow
from the $q$-algebras defined in the previous section.

As is well known the thermodynamic properties are determined by the partition
function $Z$, which in the canonical ensemble is defined by
$$
Z=Tr(e^{-\beta H})\ ,
\eqno{(3.1)}
$$
where $\beta=1/(kT)$, $H$ is the hamiltonian and the trace must be taken
over a complete set of states. For any operator $O$, the ensemble average is
then obtained with the prescription
$$<O>=\frac{1}{Z}Tr(e^{-\beta H}O)$$
and it is crucial that the cyclicity of the trace be consistent with the
algebraic structure. In the remainder of this section we restrict ourselves
to systems described by a single oscillator mode. Multimode systems are treated
in sections 4 and 5.

Two types of hamiltonians are usually dealt with in the literature. The first
one is identified with the number operator and reads as \cite{greenberg},
\cite{alth}
$$H=\omega N\ ,
\eqno{(3.2)}
$$
(hereafter $\hbar=1$), while the second one is defined in terms of the creation
and
annihilation operators, following the classical realization of the harmonic
oscillator \cite{martin}
$$H=\frac{\omega}{2}(a^+a+aa^+)\ .
\eqno{(3.3)}
$$
For the $q$-fermion oscillator the latter can be taken as
$$H=\frac{\omega}{2}(f^+f-ff^+)\ .
\eqno{(3.4)}
$$
In section 5 we shall generalize (3.3) to a general linear combination of
$a^+a$ and $aa^+$.

The partition function, and hence the thermodynamics of the system, is
determined uniquely by the spectrum of the hamiltonian. With the choice (3.2),
the partition function will be that of the harmonic oscillator,
$$Z=\sum\limits^{\infty}_{n=0}e^{-\beta\omega n}=\frac{e^{\beta\omega}}
{e^{\beta\omega}-1}
\eqno{(3.5)}
$$
irrespective of what deformation, Bose or Fermi, we want to choose. Of course,
for those special values of $q$ for which the Fock space separates into
disjoint
subspaces each carrying a finite-dimensional representation, it is perhaps
physically more meaningful to consider the truncation to one of these subspaces
only; i.e. picking only those terms in the sum (3.5) corresponding to the
subspace chosen. As an example, when $q=e^{i\pi/3}$, the Fock space states
$|3>,\ |4>,\ |5>$ span a representation $R$ of (2.1) in which
$$
a_R=\left(\begin{array}{lll}
0&i&0\\
0&0&i\\
0&0&0\end{array}\right)\ ,\  a^+_R=\left(\begin{array}{lll}
0&0&0\\
i&0&0\\
0&i&0\end{array}\right)\ ,\  N_R=\left(\begin{array}{lll}
3&0&0\\
0&4&0\\
0&0&5\end{array}\right)
\eqno{(3.6)}
$$
(note that $a^+$ is not the conjugate of $a$ in this representation). The
corresponding truncated partition function is
$$
Z_R=e^{-3\beta\omega}+e^{-4\beta\omega}+e^{-5\beta\omega}\ .
\eqno{(3.7)}
$$
Summing over all subspaces reproduces the full partition function (3.5).

While the thermodynamics for a system with hamiltonian (3.2) is independent
of the deformation, Green functions like $<a^+a>$ will depend on the
deformation. For the $q$-bosonic algebra (2.1) we obtain
$$<a^+a>=<[N]>=\frac{1}{Z}\sum\limits_n [n]e^{-\beta\omega n}=$$
$$\hspace*{0.5cm}=\frac{e^{\beta\omega}-1}{e^{2\beta\omega}-(q+q^{-1})
e^{\beta\omega}+1}\ ,
\eqno{(3.8)}
$$
while for the operators (2.5) obeying (2.6) we have the simple expression
$$<A^+A>=<[N]^B>=\frac{1}{e^{\beta\omega}-q^2}\ .
\eqno{(3.9)}
$$
Similarly, for the $q$-fermion algebras (2.11) and (2.16) we obtain
$$<f^+f>\ =\ <[N]^f>\ =\ \frac{e^{\beta\omega}-1}{e^{2\beta\omega}+(q-q^{-1})
e^{\beta\omega}-1}
\eqno{(3.10)}
$$
and
$$<F^+F>\ =\ <[N]^F>\ =\ \frac{1}{e^{\beta\omega}+q^2}\ .
\eqno{(3.11)}
$$
The average of $N$ is of course always given by the usual Bose-Einstein
formula
$$n=<N>=\frac{1}{e^{\beta\omega}-1}
\eqno{(3.12)}
$$
(when the average is taken over the complete Fock space), to which (3.8) and
(3.9) reduce in the case $q=1$, and (3.11) in the case $q=\pm\; i$, as they
should. When $q=1$ in (3.10) or (3.11), the Fermi-Dirac distribution is
recovered.

When $q\neq 1$, the temperature Green functions $<a^+a>$, etc. do not have a
direct relation to the thermodynamic quantities of the system. In a multimode
context, such as will be considered in section 4 and 5, this can be understood
by noting that the average $<a^+_ka_k>$ does not describe the average
occupancy of the $kth$ level, $<N_k>$.

When the hamiltonian is chosen as in (3.3) or (3.4), the partition function
cannot any longer be computed in closed form for generic $q$. This case has
been investigated in refs. \cite{martin}-\cite{manko}.
\vskip 1.0 cm
\section{Temperature Green function approach.}

In this section we shall define the single-particle temperature Green function
for $q$-particles satisfying the algebras (2.6) or (2.16). This would mean
that the fundamental field operators of the theory are the ones connected
to the operators appearing in (2.6) and (2.16) (see (4.4)). In the case of the
algebras (2.1) and (2.11) the construction of this function is a more
complicated task due to the nonlinearity of the commutation relations with
respect to the annihilation and creation operators, which makes it less
obvious how to define the Green function.

Let us start by defining the single-particle Green function as
$$G_{\alpha\beta}(\vec{x},\tau;\ \vec{x}',\tau')=
-<T_q\psi_\alpha(\vec{x},\tau)\psi^+_\beta(\vec{x}',\tau')>\ ,
\eqno{(4.1)}
$$
where the ordering operator $T_q$ orders the operators according to their
values of $\tau$,
$$
T_q(\psi(\vec{x},\tau)\psi^+(\vec{x}',\tau'))=\left\{\begin{array}{ll}
\psi(\vec{x},\tau)\psi^+(\vec{x}',\tau')\ {\rm\ if}\ \tau>\tau'\\
q\psi^+(\vec{x}',\tau')\psi(\vec{x},\tau)\ {\rm\ if}\ \tau<\tau'
\end{array}\right.
\eqno{(4.2)}
$$
In this section in order to simplify the notation and consider both $q$-bosonic
algebra (2.6) and $q$-fermionic algebra (2.16) simultaneously, we shall use
$q$ instead of $\pm q^2$. Note that when $q=1\ (q=-1)$ eqs. (4.1) and (4.2)
are the usual definitions for bosons (fermions). The indices $\alpha$ and
$\beta$ label the components of the field operators and in what follows they
will be omitted. Finally, the operators $\psi(\vec{x},\tau)$ and $\psi^+
(\vec{x},\tau)$ are the Heisenberg operators, related to the Schr\"odinger
ones $\psi(\vec{x})$ and $\psi^+(\bar{x})$ through the relations
$$\psi(\vec{x},\tau)=e^{H\tau}\psi(\bar{x})e^{-H\tau}\ ,$$
$$\psi^+(\vec{x},\tau)=e^{H\tau}\psi^+(\vec{x})e^{-H\tau}\ .
\eqno{(4.3)}
$$
As a simple example, we shall compute the Green function for a spin-zero
noninteracting system. In this case the states are plane waves so that
$$\psi(\vec{x})=\frac{1}{\sqrt{V}}\sum\limits_{\vec{k}}e^{i\vec{k}\vec{x}}
A_k\ ,\eqno{(4.4)}$$
$$\psi^+(\vec{x})=\frac{1}{\sqrt{V}}\sum\limits_{\vec{k}}e^{-i\vec{k}\vec{x}}
A^+_k\ ,$$
and the operators $A_k,A^+_k$ are supposed to satisfy the $q$-oscillator
algebra
$$A_kA^+_k-qA^+_kA_k=1\ ,\eqno{(4.5)}$$
$$[N_i,A_k]=-\delta_{ik}A_k,[N_i,A^+_k]=\delta_{ik}A^+_k\ .$$
The free hamiltonian will be taken as
$$H=\sum\limits_{k}\in_kN_k\ ,\eqno{(4.6)}$$
so that the following commutation relations hold
$$[H,A_k]=-\in_kA_k\ ,\ [H,A^+_k]=\in_kA^+_k\ . \eqno{(4.7)}$$
The Heisenberg field operators (4.3) are then
$$\psi(\vec{x},\tau)=\frac{1}{\sqrt{V}}\sum\limits_{\vec{k}}e^{i\vec{k}\vec{x}
-\in_k\tau}A_k\ ,
\eqno{(4.8)}$$
$$
\psi^+(\vec{x},\tau)=\frac{1}{\sqrt{V}}\sum\limits_{\bar{k}}e^{-i\vec{k}\vec{x}
+\in_k\tau}A^+_k\ .$$
Now, from the definition of the Green function (4.1) and using (4.8) we have
$$G^\circ(\vec{x},\tau;\ \vec{x}',\tau)=-\frac{1}{V}
\sum\limits_{\vec{k}\vec{k}'}e^{i(\vec{k}\vec{x}-\vec{k}'\vec{x}')}
e^{-\in_k\tau+\in_{k'}\tau'}\hspace{2.5cm}$$
$$\hspace*{3.0cm}\times\{\theta(\tau-\tau')<A_kA^+_{k'}>+\theta(\tau'-\tau)q
<A^+_{k'}A_k>\}$$
$$\hspace*{1.8cm}=-\frac{1}{V}\sum\limits_{\vec{k}}e^{i\vec{k}(\vec{x}
-\vec{x}')}e^{-\in_k(\tau-\tau')}\{\theta(\tau-\tau')<A_kA^+_k>$$
$$+\theta(\tau'-\tau)q<A^+_kA_k>\}\ .
\eqno{(4.9)}
$$
The last equality in (4.9) follows from the orthogonality of states with
different occupation numbers $\{n_k\}$. For the ensemble averages we have
according to eq. (3.9)
$$<A^+_kA_k>=\frac{1}{e^{\beta\in_k}-q}\ ,
\eqno{(4.10)}
$$
and from eq. (4.5),
$$<A_kA^+_k>=1+q<A^+_kA_k>=\frac{e^{\beta\in_k}}{e^{\beta\in_k}-q}\ .
\eqno{(4.11)}
$$
Let us now turn our attention to the "periodicity" properties of the Green
function (4.1) in the domain where $0<\tau<\beta,\ 0<\tau'<\beta$. Keeping
$\tau'$ fixed, then
$$G(\vec{x},0;\vec{x}',\tau')=-\frac{q}{Z} Tr(e^{-\beta
H}\psi^+(\vec{x}',\tau')
\psi(\vec{x},0))$$
$$\hspace*{2.4cm}=-\frac{q}{Z}Tr(\psi(\vec{x},0)e^{-\beta
H}\psi^+(\vec{x}',\tau'))$$
$$\hspace*{2.4cm}=-\frac{q}{Z}Tr(e^{-\beta
H}\psi(\vec{x},\beta)\psi^+(\vec{x}',\tau'))
$$
$$= qG(\vec{x},\beta;\vec{x}\,',\tau')\ .
\eqno{(4.12{\rm\; a})}$$
Similarly,
$$G(\vec{x},\tau;\vec{x}',\beta)=qG(\vec{x},\tau;\vec{x}',0)\ .
\eqno{(4.12{\rm\;b})}$$
Since the hamiltonian is time-independent, $G$ depends only on the difference
$\tau-\tau'$ and eqs. (4.12) may be written as
$$G(\vec{x},\vec{x}\,';\ \tau-\tau'<0)=qG(\vec{x},\vec{x}\,';\ \tau-\tau'
+\beta)\ .
\eqno{(4.13)}$$
Note that when $q=1\ (q=-1)$, $G$ is periodic (antiperiodic) with period
$\beta$ in the range $0\leq\tau,\ \tau'\leq\beta$, which corresponds to the
usual bosonic (fermionic) case.

Now the next step is to obtain an appropriate Fourier decomposition in the
variable $\tau$ for the Green function satisfying the boundary conditions
(4.13). For this purpose, we introduce a new function
$$g(\vec{x},\vec{x}\,',\tau)=q^{\tau/\beta}G(\vec{x},\vec{x}\,',\tau)\ ,
\eqno{(4.14)}
$$
which according to (4.13), is periodic in $\tau$ variable with period $\beta$.
Thus this function can be expanded in a Fourier series
$$g(\vec{x},\vec{x}',\tau)=\frac{1}{\beta}\sum\limits_{n}e^{-i\omega_n\tau}
g(\vec{x},\vec{x}',\omega_n)\ ,
\eqno{(4.15)}$$
where $\omega_n=2n\pi/\beta$. The associated Fourier coefficient is then
given by
$$g(\vec{x},\vec{x}',\omega_n)=\int\limits^\beta_0 d\tau\ e^{i\omega_n\tau}
g(\vec{x},\vec{x}\,',\tau)\ .
\eqno{(4.16)}
$$
Eqs. (4.15) and (4.16) can be written in terms of $G$,
$$G(\vec{x},\vec{x}\,',\tau)=\frac{1}{\beta}\sum\limits_n e^{-ik_4\tau}
g(\vec{x},\vec{x}\,',k_4)\ ,
\eqno{(4.17)}$$
$$g(\vec{x},\vec{x}',k_4)=\int\limits^\beta_0 d\tau\ e^{ik_4\tau}G(\vec{x},
\vec{x}\,',\tau)\ ,
\eqno{(4.18)}
$$
where
$$k_4=2n\pi/\beta -\frac{i}{\beta}\ln q\ .
\eqno{(4.19)}
$$
When $q=1\ (q=-1)$, then $k_4=2n\pi/\beta\ (k_4=(2n+1)\pi/\beta)$, i.e., we
have the common expressions for the bosonic (fermionic) frequencies.

For the noninteracting Green function (4.9) we have then
$$g^\circ(\vec{x}-\vec{x}',k_4)=\int\limits^\beta_0 d\tau\ e^{ik_4\tau}
G^\circ(\vec{x}-\vec{x}',\tau)\hspace{4.7cm}$$
$$\hspace*{1.0cm} =-\frac{1}{V}\sum\limits_{\vec{k}}e^{i\vec{k}(\vec{x}-
\vec{x}')}
\int\limits^\beta_0 d\tau\ e^{(ik_4-\in_k)\tau}<A_kA^+_k>$$
$$\hspace*{2.9cm}=\frac{1}{V}\sum\limits_{\vec{k}}\ \frac{e^{i\vec{k}
(\vec{x}-\vec{x}')}}
{ik_4-\in_k}\left\{\left(-e^{(ik_4-\in_k)\beta}+1\right)<A_kA^+_k>\right\}\ .
\eqno{(4.20)}
$$
The condition (4.19) and eq. (4.11) imply that the expression in brackets
$\{...\}$ in (4.20) is equal to 1 and consequently, we have
$$g^\circ(\vec{x}-\vec{x}\,',k_4)=\frac{1}{V}\sum\limits_{\vec{k}}
\frac{e^{i\vec{k}(\vec{x}-\vec{x}')}}{ik_4-\in_k}\ ,$$
or after performing the Fourier decomposition in the space variables,
$$g^\circ(\vec{k},k_4)=\frac{1}{ik_4-\in_k}\ .
\eqno{(4.21)}
$$
We see that the free $q$-propagator has the usual form with the only
difference that now the frequencies are given by eq. (4.19).

To conclude this section let us discuss the method for evaluating frequency
sums appearing in the calculation of statistical averages. Suppose that for a
function $F(...,k_4)$ we want to evaluate the sum $\frac{1}{\beta}\sum\limits_n
F(...,k_4)$. By choosing a function having poles at the values $z=k_4$ and
with unit residue we can reduce the summation over $k_4$ to evaluation of a
contour integral. One possible choice are the functions
$$f^\pm_q(z)=\frac{\pm i\beta}{1-q^{\pm 1}\ e^{\mp i\beta z}}\ .
\eqno{(4.22)}
$$
Then we have
$$\frac{1}{\beta}\sum\limits_n F(...,k_4)=-\frac{1}{\beta}\sum\limits_jf^\pm_q
(z_j)\ Res\ F(...,z_j)\ ,
\eqno{(4.23)}
$$
where $z_j$ are the poles of $F$ in the complex plane defined by the variable
$z=k_4$. Let us remark also that the choice of $f^+_q$ or $f^-_q$ is determined
by the requirement that the integral $\oint fFdz$ should converge on
the circumference of an infinite-radius circle in the complex plane.

As an example of the use of formula (4.23), let us calculate again the thermal
average $<A^+_kA_k>$ but this time by means of the Green function.

{}From the definition (4.1) and the expansions (4.8), it is easy to show that
$$<A^+_kA_k>=-\frac{1}{q}\;^{lim}_{\epsilon\rightarrow 0}\ \frac{1}{\beta}
\sum\limits_n\frac{e^{ik_4\epsilon}}{ik_4-\in_k}\ ,
\eqno{(4.24)}
$$
where $\epsilon>0$. (The factor $e^{ik_4\epsilon}$ is a convergence factor
and therefore $\epsilon$ must be kept until the sum is evaluated). Using the
formula (4.23) with the function $f^-_q$ defined in (4.22), we get
$$<A^+_kA_k>=\frac{1}{q\beta}\ \frac{-i\beta}{1-q^{-1}e^{i\beta z}}\
\frac{1}{i}
\left|_{z=-i\in_k}\right.$$
$$=\frac{1}{e^{\beta\in_k}-q}\ ,$$
i.e., we reproduce the same result obtained in the previous section (see eqs.
(3.9) and (3.11)) from statistical mechanics.

To develop an analogous Green function method for anyonic fields is of
considerable interest. This question is under study.
\vskip 1.0 cm
\section{The ideal $q$-gas}

By an "ideal $q$-gas" we understand a system defined by the hamiltonian
$$H=\sum\limits_i\epsilon_i\ (\alpha\ a^+_ia_i\ +\ (1-\alpha)\ a_ia^+_i)\ ,
\eqno{(5.1)}
$$
where the operators $a_i,a^+_i$ satisfy the $q$-oscillator algebra (2.21),
and $\alpha$ is a real parameter between 0 and 1. An equivalent form of the
hamiltonian can be written using (2.4):
$$H=\sum\limits_i\epsilon_i\ (\alpha[N_i]\ +\ (1-\alpha)[N_i+1])\ .
\eqno{(5.2)}
$$
We shall interpret $a_i,a^+_i,N_i$ as annihilation, creation and occupation
number operators, respectively, of particles in the state (level) $i$,
although the mathematical results are, of course, independent of this
interpretation. We shall cal $\epsilon_i$ the energy of the level $i$. The
ideal $q$-gas is a special case of a general class of systems, which could be
called "general ideal gases". The energy eigenvalues of such a system are
uniquely determined by a set of occupation numbers $\{n_i\}$ and a set of
functions $E_i$:
$$E=\sum\limits_i\ E_i(n_i)
\eqno{(5.3)}
$$
In a usual ideal gas, the functions $E_i$ are proportional to the occupation
number:
$$E_i(n)=\epsilon^\circ_in\ ,
\eqno{(5.4)}
$$
corresponding to the physical interpretation of the total energy as a sum
of single-particle energies $\epsilon^\circ_i$, with vanishing interaction
energy between the particles. In the general case (5.3) the interaction between
the particles is such that it causes the energy of the ith level to depend
on the occupancy of that level. For the ideal $q$-gas (5.2), this
dependence is for real $q=\exp t$
$$E_i(n)=\frac{\epsilon_i}{{\rm sinh}t}(\alpha\;{\rm sinh}\,(nt)+(1-\alpha)
{\rm sinh}\,((n+1)t)),
\eqno{(5.5)}
$$
and for $q=\exp (i\theta)$
$$E_i(n)=\frac{\epsilon_i}{\sin\theta}(\alpha\sin (n\theta)+(1-\alpha)\sin
((n+1)\theta))\ .
\eqno{(5.6)}
$$
(In an interpretation, where $a_i,a^+_i$ and $N_i$ are the ladder operators
and excitation levels of oscillators, (5.5) and (5.6) correspond to a
certain type of nonlinear oscillator \cite{maniko}).

The grand canonical partition function
$$Z=Tr\exp (-\beta(H-\mu N))=\exp (-\beta\Omega)\ ,
\eqno{(5.7)}
$$
where $N$ is the total number operator
$$N=\sum\limits_i N_i\ ,
\eqno{(5.8)}
$$
$\mu$ the corresponding chemical potential, and $\Omega$ the grand canonical
potential, factorizes for a general ideal gas into a product of single level
partition functions:
$$Z=\prod\limits_i Z_1(i,\beta,\mu)\ ,
\eqno{(5.9)}
$$
$$Z_1(i,\beta,\mu)=\sum\limits^\infty_{n=0} e^{-\beta(E_i(n)-\mu n)}\ .
\eqno{(5.10)}
$$
Correspondingly, the grand canonical potential is given as a sum over the
levels $i$:
$$\Omega=-\frac{1}{\beta}\sum\limits_i \log Z_1(i,\beta,\mu)\ .
\eqno{(5.11)}
$$
The allowed values of the chemical potential $\mu$ are determined by the
requirement that the sum in (5.10) converges. For the ideal $q$-gas with
real $q$, the absolute value of $E_i$ grows exponentially with $n$, eq. (5.5).
In order to have the total energy bounded from below, we require $\epsilon_i
\geq 0$ for all $i$ in (5.5). Then $\mu$ can take any real value if all
$\epsilon_i >0$, if at least one $\epsilon_i=0$, we must have $\mu <0$.
When $q$ is a pure phase, $|E_i(n)|$ stays bounded for all in (eq. (5.6)),
in this case we require $\mu <0$.

The sum over $n$ in (5.10) cannot be explicitly performed in general.
However, when $q=\exp t$ is close to 1 ($t\rightarrow 0$), we can calculate
the leading correction to the usual Bose gas. To order $t^2$, eq. (5.5)
reads
$$E_i=\epsilon_i(n+1-\alpha +\frac{1}{6}n(n+1)(n+2-3\alpha)t^2+\cdots)\ .
\eqno{(5.12)}
$$
Substituting this into (5.10) gives the approximate expression
$$Z_1(i,\beta,\mu)\simeq e^{\beta(\alpha-1)\epsilon_i}\sum\limits^\infty_{n=0}
e^{\beta(\mu-\epsilon_i)n}(1-\frac{t^2}{6}\beta\epsilon_i n(n+1)(n+2-3\alpha)
+\cdots)\ .
\eqno{(5.13)}
$$
Introducing the variable $x_i=\beta\epsilon_i$ and the fugacity $z=\exp (\beta
\mu)$ and performing the sum over $n$ we obtain
$$Z_1(x_i,Z)\simeq e^{(\alpha-1)x_i}Z_0(z,x_i)(1-t^2x_ize^{-x_i}Z^3_0(z,x_i)
(\alpha ze^{-x_i}+1-\alpha))\ ,
\eqno{(5.14)}
$$
where
$$Z_0(z,x)=\frac{1}{1-ze^{-x}}$$
corresponds to the ideal Bose gas. The grand canonical potential is then to
second order in $t$
$$\beta\Omega\simeq\sum\limits_i(1-\alpha)x_i-\sum\limits_i\log Z_0(z,x_i)
\eqno{(5.15)}
$$
$$+t^2z\sum\limits_ix_ie^{-x_i}Z^3_0(z,x_i)(\alpha ze^{-x_i}+1-\alpha)\ .
$$
In order to study the thermodynamics of the system we have to specify the
energies $\epsilon_i$ and the density of states. We shall consider a system
where the states $i$ are specified by a d-dimensional momentum vector $\vec{k}$
(the momentum of the $q$-boson). The energy of the $q$-boson is given by the
dispersion law
$$\epsilon_i=\epsilon(\vec{k})=\gamma |\vec{k}|^p\ ,
\eqno{(5.16)}
$$
covering the cases of nonrelativistic $(\gamma=1/2 m,\ p=2)$ and
ultrarelativistic $(\gamma=1,\ p=1)$ $q$-bosons. We enclose the system in a
large d-dimensional volume $V$ and replace in the usual way the sum over
levels by an integral over $\vec{k}$-space:
$$\sum\limits_i\rightarrow V\int \frac{d^dk}{(2\pi)^d}\ .$$
The first term on the right hand side of (5.15) is now divergent, giving rise
to an infinite (negative) vacuum pressure, and we renormalize it away by
simply dropping it. This means in effect that we subtract away the vacuum
energy, i.e. replace the functions $E_i(n)$ by $E_i(n)-E_i(0)$.
\vskip 0.5 cm
Performing the $\vec{k}$-space integrals using the general formula
$$\int^\infty_0 dx\ \frac{x^{\alpha-1}e^{-px}}{(e^{qx}-z)^n}\ =
\frac{\Gamma(\alpha)}{(n-1)!}\sum\limits^\infty_{k=0}\frac{(k+n-1)!}{k!}
\frac{z^k}{(qk+qn+p)^\alpha}
\eqno{(5.17)}
$$
and introducing the functions $g_\ell(z)$
$$g_\ell(z)=\sum\limits^\infty_{k=1}\ \frac{z^k}{k^\ell}\ ,
\eqno{(5.18)}
$$
we get for the pressure $p=-\Omega/V$
$$\beta p=a\left\{g_{\frac{d}{p}+1}(z)-\frac{d}{2p}\ t^2\left[g_{\frac{d}{p}
-1}(z) +(1-2\alpha)g_{\frac{d}{p}}(z)\right]\right\}\ .
\eqno{(5.19)}
$$
Here
$$a=\frac{A_{d-1}}{(2\pi)^d}\ \frac{\Gamma(\frac{d}{p}+1)}{d}\ \left(\frac{1}
{\gamma\beta}\right)^{d/p}\ ,
\eqno{(5.20)}
$$
where $A_{d-1}$ is the area of the unit sphere $S^{d-1}$:
$$A_{d-1}=\frac{2\pi^{d/2}}{\Gamma(\frac{d}{2})}\ .$$
The $q$-boson density
$$n=\frac{N}{V}=\left(\frac{\partial p}{\partial\mu}\right)_{T,V}=z\left(
\frac{\partial(\beta p)}{\partial z}\right)_\beta
\eqno{(5.21)}
$$
is easily found, since
$$z\frac{dg_\ell(z)}{dz}=g_{\ell-1}(z)\ .$$
We get
$$n=a\left\{g_{\frac{d}{p}}(z)-\frac{d}{2p}\ t^2\left[g_{\frac{d}{p}-2}(z)+
(1-2\alpha)g_{\frac{d}{p}-1}(z)\right]\right\}\ .
\eqno{(5.22)}
$$
According to our general considerations, the chemical potential is restricted
to negative values for the system we consider.\\
Since
$$g_\ell(z)>g_{\ell'}(z)$$
when $\ell <\ell'$ and $z\rightarrow 1$, eq. (5.19) implies that the pressure
becomes negative for $\mu$ sufficiently close to 0 ($z$ sufficiently close
to 1) if $t$ is real. This is not a signal of instability of the ideal $q$-gas,
rather it signals the breakdown of the expansion in $t$. From (5.13) it is
clear that the effective expansion parameter is $t^2$ times a positive power
of $Z_0(z,x_i)$, and this grows large when $z\rightarrow 1$ and $x_i\rightarrow
0$. Although the problem of negative pressure is absent for imaginary $t$,
also in this case the expansion cannot be trusted as $z\rightarrow 1$. Thus
we are not able to address the interesting question how the phenomenon of
Bose-Einstein condensation is affected by the deformation.

For low densities, however, the expansion in $t^2$ is meaningful. We invert
eq. (5.22) to obtain $z$ as a series in $n$:
$$z=a_1(\frac{n}{a})+a_2(\frac{n}{a})^2+\cdots\ ,
\eqno{(5.23)}
$$
where
$$a_1=1+\frac{d}{p}(1-\alpha)t^2\ ,$$
$$a^2=-\frac{1}{2^{d/p}}(1-\frac{d}{p}\alpha t^2)\ .$$
Substituting (5.23) into (5.19) and expanding in powers of $n$, we obtain
the virial expansion of the equation of state
$$\beta p=n(1+Bn+\cdots)\ .
\eqno{(5.24)}
$$
The second virial coefficient is given by
$$B=-\frac{1}{2^{d/p+1}a}(1-\frac{d}{p}t^2)\ .
\eqno{(5.25)}
$$
Thus, for very low densities, the $q$-gas behaves as a classical ideal gas.
The expression for the second virial coefficient is independent of the
parameter $\alpha$ in the hamiltonian, and shows that
the deformation weakens the attraction between pairs of $q$-bosons
producing an increase in the pressure, when $t$ is real, whereas the effect
of the deformation is the opposite, when $t$ is imaginary.

Our results in this section generalize previous partial results obtained e.g.
in refs. \cite{martin} and \cite{maniko}. The papers in ref. \cite{molin}
address the same questions as we have done in this section using the
retarded and advanced Green functions. Their results, where {\bf both} the
chemical potential $\mu$ {\bf and} the particle number $N$ enter into the
average distribution number and have sharp values simultaneously are
meaningless, however.
\vskip 1.0 cm
\section{Anyonic gas near Bose and Fermi statistics.}

Some of the statistical properties of the anyonic gas have been studied in
certain approximations \cite{schrie},\cite{comte}. In this section we will
find perturbatively (in the first order of the statistical parameter) the
equation of state for a gas of anyons near Bose and Fermi statistics by using
instead the temperature Green function method. The same results have been
obtained in \cite{comte} in a different approach.

We consider a system of $N$ identical anyons described by the hamiltonian
\cite{wile}
$$H=\sum\limits^N_{i=1}\ \frac{(\vec{p}_i-\vec{A}_i)^2}{2m}\ ,
\eqno{(6.1)}
$$
where
$$\vec{A}_i=\alpha\sum\limits_{i\neq j}\frac{\vec{k}\times\vec{r}_{ij}}
{\vec{r}\,^2_{ij}}\ ,\ (\vec{r}_{ij}\equiv\vec{r}_i-\vec{r}_j)\ ,
\eqno{(6.2)}
$$
is the statistical gauge field, $\vec{k}$ is a unit vector perpendicular to
the plane and $\alpha=e\phi/\pi$ is the statistical parameter ($e$ and $\phi$
are the charge and the flux carried by each anyon).

Due to the singular $\alpha^2$-interaction term in (6.1), the standard
perturbation
treatment requires a redefinition of the $N$-body anyonic wave function, so
that the latter vanishes when any two anyons approach each other \cite{comte}.
Therefore, the real wave function is defined as
$$\psi_p(\vec{r}_1,...,\vec{r}_N)=\prod\limits_{i<j}\ r^{|\alpha|}_{ij}
\tilde{\psi}_p(\vec{r}_i,...,\vec{r}_N)$$
For the new function $\tilde{\psi}_p$ the hamiltonian reads as
$$\tilde{H}=\sum\limits^N_{i=1}\left\{\frac{\vec{p}\,^2_i}{2m}+
\frac{i\alpha}{m}
\sum\limits_{j\neq i}\ \frac{\vec{k}\times\vec{r}_{ij}}{r^2_{ij}}\
\vec{\partial}
_i-\frac{|\alpha|}{m}\sum\limits_{j\neq i}\ \frac{\vec{r}_{ij}}{r^2_{ij}}\
\vec{\partial}_i\right\}
\eqno{(6.3)}
$$
and consequently, the singular terms diverging like $1/r^2_{ij}$ are
cancelled. Now the perturbative analysis can be carried out. At order
$\alpha$, only the term proportional to $|\alpha|$ in (6.3) contributes to the
grand partition function. Indeed, the contribution of the second term
proportional to $\alpha$ can be shown to vanish due to the symmetry of the free
spectrum under the change $p_x\leftrightarrow p_y$ (i.e., the corrections
to the spectrum coming from the $\alpha$ interaction cancel in the order
$\alpha$ in the grand partition function). Furthermore, the $|\alpha|$
interaction term in (6.3) can be replaced (in a perturbative sense) by a sum
of two-body $\delta$-potentials \cite{comte}
$$V=\frac{2\pi|\alpha|}{m}\sum\limits_{i<j}\delta^2(\vec{r}_i-\vec{r}_j)\ .
\eqno{(6.4)}
$$
This fact will allow us to simplify the calculations of the self-energy
corrections.

Let us consider now the statistics. As is well known the temperature Green
function satisfies the Dyson equation, which in the momentum space has the
form \cite{fetter}
$$G(\vec{k},\omega_n)=G^\circ(\vec{k},\omega_n)+G^\circ(\vec{k},\omega_n)
\sum(\vec{k},\omega_n)G(\vec{k},\omega_n)\ ,
\eqno{(6.5)}
$$
where $\sum$ is the proper self-energy and $G^\circ$ is the free propagator,
$$G^\circ=\frac{1}{i\omega_n-(\in^\circ_k-\mu)}\ ,
\eqno{(6.6)}
$$
$\omega_n=2n\pi/\beta\ (\omega_n=(2n+1)\pi/\beta)$ for bosons (fermions),
$\in^\circ_k=\vec{k}^2/(2m)$ and $\mu$ is the chemical potential. Eq. (6.5)
has the explicit solution
$$
G(\vec{k},\omega_n)=\frac{1}{i\omega_n-(\in^\circ_k-\mu)-\sum(\vec{k},\omega_n)}
\ .
\eqno{(6.7)}
$$
The first-order self-energy is easily evaluated and is given by \cite{fetter}
$$\Sigma_{(1)}(\vec{k}_1\omega_n)\equiv\Sigma_{(1)}(\vec{k})=V(0)\int
\frac{d^2k}{(2\pi)^2}n^\circ_k$$
$$\hspace*{0.5cm}\pm \int\frac{d^2k'}{(2\pi)^2}V(\vec{k}-\vec{k}')n^\circ_{k'}\
,
\eqno{(6.8)}
$$
where the upper (lower) sign corresponds to bosons (fermions),
$$n^\circ_k=\frac{1}{e^{\beta(\in^0_k-\mu)}\mp 1}
\eqno{(6.9)}
$$
is the usual Bose (Fermi) distribution and $V$ is the two-particle interaction
potential.

According to (6.4), for anyons we have
$$V(\vec{k})=\frac{2\pi}{m}|\alpha|
\eqno{(6.10)}
$$
and therefore, $\sum_{(1)}$ is easily evaluated. Finally we obtain for bosons
$$\Sigma^\beta_{(1)}=\frac{2|\alpha|}{\beta}\int\limits^\infty_0 dx\ \frac{z}
{e^x-z}=-\frac{2|\alpha|}{\beta}\ln (1-z)\ ,
\eqno{(6.11)}
$$
while for fermions,
$$\Sigma^F_{(1)}=0\ .
\eqno{(6.12)}
$$
We conclude that there are no connections of order $\alpha$ to the
thermodynamical quantities of the anyonic gas near Fermi statistics.

Since $\sum^B_{(1)}$ is independent of $\omega_n$, the particle distribution
will be given by the same expression as the unperturbed one (6.9) with the
only difference that instead of the free energy $\in^\circ_k$, we will have
$\in^\circ_k+\sum^B_{(1)}$. Besides, in the approximation we are dealing with
(first order in $\alpha$), $\sum^B_{(1)}$ is also independent of $\vec{k}$
and thus, the density of particles is easily found,
$$n=\int\frac{d^2k}{(2\pi)^2}n_k=-\frac{1}{\lambda^2}\ln (1-ze^{-\beta
\sum^B_{(1)}})$$
$$\sim -\frac{1}{\lambda^2}\ln (1-z+z\beta\Sigma^B_{(1)})\sim
-\frac{1}{\lambda^2}
\left\{\ln (1-z)+\frac{z\beta\sum^B_{(1)}}{1-z}\right\}$$
where $\lambda=\sqrt{\frac{2\pi\beta}{m}}$\ .

Substituting eq. (6.11) into the latter equation, we obtain
$$n=-\frac{1}{\lambda^2}\ln (1-z)\left\{ 1-\frac{2|\alpha|z}{1-z}\right\}\ .
\eqno{(6.13)}
$$
For the pressure $p$, in turn, we find
$$p\beta=\int
dz\frac{n}{z}=\frac{1}{\lambda^2}\left\{\sum\limits^\infty_{\ell=1}
\frac{z^\ell}{\ell^2}-|\alpha|\ln^2((1-z)\right\}\ .
\eqno{(6.14)}
$$
Eqs. (6.13) and (6.14) were found in \cite{comte} by following a different
procedure. These equations can be rewritten in terms of the functions
$g_\ell(z)$ defined in (5-18). We have then
$$n=\frac{1}{\lambda^2}g_1(z)\{1-2|\alpha|g_0(z)\}\ ,
\eqno{(6.15)}
$$
$$p\beta=\frac{1}{\lambda^2}\{g_2(z)-|\alpha|g^2_1(z)\}\ .
\eqno{(6.16)}
$$
Expanding $p$ in powers of $n$ we get the following expression for the second
virial coefficient,
$$B=-\frac{\lambda^2}{4}+|\alpha|\lambda^2\ .
\eqno{(6.17)}
$$
This expression shows that $B$ exhibits a nonanalytic behaviour (a cusp at
$\alpha=0$) as a function of the statistics determining parameter $\alpha$.

Comparing eqs. (5.19), (5.22) and (5.25) for the two-dimensional
non-relativistic
$q$-gas with eqs. (6.15), (6.16) and (6.17) for the anyonic gas, we notice
that they differ from each other and consequently, one sees explicitly that
these two objects indeed describe different physical systems.
\vskip 1.0 cm
\section{Conclusions.}

In this paper we have studied some statistical properties of $q$-oscillators,
quons and anyons in order to shed some light on the interrelation as well
as on the basic differences among these objects. In particular, the statistics
of bosonic and fermionic $q$-oscillators have been considered in detail for
different Fock space representations of the corresponding algebras in the
case when the hamiltonian is identified with the number operator. The choice
of this hamiltonian allowed us to calculate the statistical averages and also
to introduce the Heisenberg picture of operators. The Green function method
has been also introduced for nonrelativistic $q$-particles by assuming that
Green function satisfies $q$-deformed periodicity conditions. This general
method is hoped to be employed in attacking the problem of formulating the
statistics of an anyonic gas.
\vskip 1.0 cm
{\bf Acknowledgements}

\noindent One of us (R.G.F.) would like to thank the World Laboratory for
financial support which enabled him to carry out this work.
\pagebreak

\end{document}